\documentclass[12pt, onecolumn, draftclsnofoot]{IEEEtran}

\usepackage{cite}
\usepackage{amsmath,amssymb,amsfonts}
\usepackage{algorithm}
\usepackage{algorithmic}
\usepackage{graphicx}
\usepackage{textcomp}
\usepackage{xcolor}
\usepackage{booktabs}

\def\BibTeX{{\rm B\kern-.05em{\sc i\kern-.025em b}\kern-.08em
    T\kern-.1667em\lower.7ex\hbox{E}\kern-.125emX}}

\usepackage[letterpaper, top=1in, bottom=1in, left=1in, right=1in]{geometry}

\begin{document}

\title{\vspace*{0.5cm} Lightweight Trustworthy Distributed Clustering}

\author{
\IEEEauthorblockN{
Hongyang Li\IEEEauthorrefmark{1},
Caesar Wu\IEEEauthorrefmark{1},
Mohammed Chadli\IEEEauthorrefmark{2},\\
Said Mammar\IEEEauthorrefmark{2},
Pascal Bouvry\IEEEauthorrefmark{1}
}

\IEEEauthorblockA{\IEEEauthorrefmark{1}University of Luxembourg\\
\texttt{\{hongyang.li, caesar.wu, pascal.bouvry\}@uni.lu}}

\IEEEauthorblockA{\IEEEauthorrefmark{2}University Paris-Saclay\\
\texttt{\{said.mammar, mohammed.chadli\}@univ-evry.fr}}
}

\maketitle

\begin{abstract}

Ensuring data trustworthiness within individual edge nodes while facilitating collaborative data processing poses a critical challenge in edge computing systems (ECS), particularly in resource-constrained scenarios such as autonomous systems sensor networks, industrial IoT, and smart cities. This paper presents a lightweight, fully distributed \(k\)-means clustering algorithm specifically adapted for edge environments, leveraging a distributed averaging approach with additive secret sharing, a secure multiparty computation technique, during the cluster center update phase to ensure the accuracy and trustworthiness of data across nodes.

\end{abstract}


\section{Introduction}
\label{Sect-Intro}
%what is the topic:

Edge computing, a paradigm emerging from distributed computing, emphasizes processing data at or near its source to minimize latency and reduce bandwidth consumption \cite{khan2019edge, li2023adaptive, li2024trustworthiness}. The rapid advancements in edge computing technologies, including algorithms for decentralized and efficient data processing, have significantly accelerated the deployment of distributed sensor networks. These networks, consisting of numerous spatially distributed sensor nodes, facilitate extensive data collection and enable real-time processing, particularly in dynamic and latency-sensitive environments \cite{shi2016edge, mao2017survey}.

Two key properties of ECS are crucial in large-scale deployments. First, ECS often employ a distributed architecture \cite{mosk2010fully}, where data processing and communication are localized or involve collaboration among a subset of edge nodes, reducing reliance on centralized servers. This decentralized structure enables excellent scalability, as additional edge devices can integrate into the network with minimal dependencies, resembling the functionality of a randomly connected peer-to-peer (P2P) network \cite{azimi2017distributed}. Second, ECS frequently operate under strict resource constraints, including limited computational power, bandwidth, and energy availability, often due to the hardware's cost-efficiency requirements and latency-sensitive applications \cite{jiang2020energy}. To address these challenges, lightweight and scalable distributed algorithms are essential to optimize resource utilization and ensure efficient operations in large-scale edge deployments.

Among existing clustering algorithms, the $k$-means algorithm \cite{ pei2025adaptive, huang2024near, hamerly2003learning} is particularly well-suited to resource-constrained environments due to its lightweight nature, balancing simplicity with computational efficiency. Unlike more sophisticated methods, $k$-means involves simple iterative updates with low computational complexity, which makes $k$-means an ideal lightweight clustering solution for edge computing systems, where maintaining low latency and resource efficiency is critical.

%These attributes make \(k\)-means clustering uniquely qualified for deployment in scenarios where rapid, reliable data processing is essential. 

Specifically, the \(k\)-means algorithm is implemented through two primary steps: initially, the cluster assignment step, in which each data point is allocated to a cluster according to its distance from the cluster centers, and subsequently, the cluster center updating step, where the cluster centers are recalculated based on these new assignments.

However, the collaborative framework of distributed \(k\)-means clustering in sensor networks, while effective for certain tasks, may introduce significant vulnerabilities, especially in adversarial environments. When data from numerous sensor nodes is shared and aggregated, the potential for data leakage, tampering, and adversarial attacks grows, underscoring the need to protect sensitive information and maintain reliable system performance \cite{mcmahan2017communication, agrawal2000privacy}. For distributed \(k\)-means clustering, the cluster assignment step typically does not present trustworthy issues, as it is performed locally within each node. In contrast, the cluster center updating step in an ECS necessitates a data aggregation protocol, such as those found in average consensus algorithms \cite{johansson2008faster}, gossip algorithms \cite{boyd2006randomized,dimakis2010gossip}. This step raises trustworthy concerns because these aggregation protocols require nodes to exchange information, which unavoidably exposes the sensitive data stored on each node. This interconnectedness heightens trust-related challenges, as a single node’s malicious activity or compromised data can disrupt the entire network. Achieving trustworthiness demands ensuring data confidentiality, maintaining accuracy, and enhancing robustness against adversarial threats—key objectives that lie at the core of our research.

In this paper, we propose a lightweight, trustworthy distributed 
\(k\)-means clustering algorithm specifically designed for ECS, such as sensor networks. Our approach not only addresses the computational and communication constraints of these networks but also ensures trustworthiness and accuracy performance of system, enabling secure and cost-effective data processing in real-time environments.

\section{Related Work}

Recent studies have addressed various challenges in sensor networks, particularly in resource-constrained environments \cite{egwuche2023machine, rao2023using, bangotra2022trust, sureshkumar2020fuzzy, lv2023unmanned, feng2023age}. These works highlight the importance of collaborative tasks within ECS.

Most existing trustworthy \(k\)-means clustering approaches, such as \cite{samet2007privacy,su2007privacy,bunn2007secure,doganay2008distributed,xing2017mutual,fan2021ppmck,zhao2021k, li2024privacy}, however, are not designed for fully distributed networks. Instead, they typically focus on data trustworthiness in server-centric settings \cite{sakuma2010large}, where multiple servers collect data from users and each server aims to protect its data from other servers. Techniques from secure multiparty computation, such as those in \cite{goldreich2009foundations}, are commonly applied in these settings, as they enable nodes to jointly compute a function while keeping their inputs secure. A comprehensive overview of these approaches can be found in \cite{meskine2012privacy}.

Applying these algorithms directly to ECS, however, presents challenges, as they generally do not scale well with an increasing number of nodes and often assume only two-party settings. A more user-centric approach \cite{sakuma2010large}, which combines Paillier cryptosystems \cite{paillier1999public} and secure function evaluation (SFE) \cite{goldreich2009foundations,yao1986generate} with a random gossip algorithm \cite{boyd2006randomized}, offers better scalability and could theoretically be applied to ECS. This method securely computes cluster centers and assigns cluster labels in each \(k\)-means iteration, making it a promising candidate for distributed networks.

However, this approach is not practically applicable in sensor networks due to its high computational complexity and significant communication bandwidth requirements, which are particularly problematic in such resource-constrained environments. As shown in \cite{meskine2012privacy}, the additive secret sharing scheme is a more promising alternative when compared to Yao’s circuit evaluation \cite{yao1986generate} and Paillier cryptosystems \cite{paillier1999public}, especially regarding computational efficiency. The key idea behind additive secret sharing is to break a secret into smaller shares, with the reconstruction relying on simple addition operations. This makes it more suitable for distributed sensor networks, where low-latency communication and minimal computational overhead are critical.

Therefore, we apply the additive secret sharing scheme in a fully distributed setting specifically designed for sensor networks. This method enables secure execution of k-means clustering while ensuring that the private data, such as the sensor measurements from individual nodes and their corresponding cluster labels, remain protected. This approach addresses the need for both trustworthiness and efficient communication in sensor networks.

%In this paper, we propose a lightweight solution for privacy-preserving distributed k-means clustering algorithm in WSNs. 

%By applying additive secret sharing, the concerned private information is divided into a number of so-called shares and it can be reconstructed only if all the shares are collected. 

%We also prove that the proposed approach does not affect the performance of the clustering algorithm. The theoretical analysis shows that the maximum information leakage is upper bounded and the desired individual privacy is protected under the condition of honest majority. 

\section{Preliminaries and Problem Definition}

\subsection{Distributed \(k\)-Means Clustering over ECS}\label{kmeansOri}

A ECS can be modeled as an undirected graph \( G = (V, E) \), with \( V = \{1, 2, \dots, n\} \) representing the set of nodes and \( E \subseteq V \times V \) denoting the set of edges. Each node \( i \in V \) can communicate only with the nodes within its neighborhood, denoted by \( d_i = \{j \mid (i, j) \in E\} \), which defines a fully distributed setting. Let \( x_i \) represent the observation data at node \( i \), and let \( X = [x_1, x_2, \dots, x_n] \) denote the entire dataset across the network. The objective of \(k\)-means clustering is to partition \( X \) into \( k \) clusters \( K = \{1, 2, \dots, k\} \), with each cluster associated with a central point, \( c_j \) for \( j \in K \). The two steps in the \(k\)-means algorithm at iteration \( t \) are :

1) \textbf{Cluster Assignment Step}

Each node \( i \) determines its closest cluster by calculating:
\begin{align}\label{clusterLabel}
l_i(t) = \underset{j \in K}{\arg\min} \, \big\| x_i - c_j(t) \big\|^2,
\end{align}

where \( \big\| x_i - c_j(t) \big\|^2 \) represents the squared Euclidean distance between data point \( x_i \) and cluster center \( c_j(t) \).

2) \textbf{Cluster Center Updating Step}

Each cluster center \( c_j \) is updated by averaging the data points assigned to it:
\begin{align}\label{clustercenter}
c_j(t+1) = \frac{1}{N_j} \sum_{x_i \in \mathcal{C}_{j(t)}} x_i,
\end{align}

where \( \mathcal{C}_{j(t)} = \big\{ x_i \mid l_i(t) = j \big\} \) denotes the data points assigned to cluster \( j \) at iteration \( t \), and \( N_j = \big| \mathcal{C}_{j(t)} \big| \) is the number of points in that cluster.

In the fully distributed context, the cluster center updating step generally requires a distributed data aggregation protocol \cite{johansson2008faster, boyd2006randomized, dimakis2010gossip, boyd2011distributed, zhang2018distributed}, enabling each node to iteratively exchange information only with its neighboring nodes to collectively compute the cluster centers. For simplicity, we assume each observation \( x_i \) is a scalar, though this approach can be extended to higher dimensions.

\subsection{Trustworthy Concerns and Adversary Model}
Trustworthy operation in  ECS relies on clearly defining what qualifies as sensitive data. Each node, \(x_i\), typically regards its observed data as private, even when collaborating on shared tasks like network localization. Participation in such tasks does not imply that a node consents to reveal its individual measurements. For example, while contributing to network localization, a node's specific location should remain undisclosed. Similarly, a node's assigned cluster label, \(l_i^f\), must also be kept confidential, as it could inadvertently expose sensitive details. On the other hand, cluster centers are not treated as private because they represent aggregated information about the group rather than individual nodes.

An essential step in designing reliable algorithms is defining a clear adversarial model. In this work, we consider the honest-but-curious model, sometimes referred to as the semi-honest or passive model~\cite{Cramer2015}. In this framework, nodes execute the algorithm as intended but may attempt to uncover private details about others. These curious nodes, while adhering to protocol, can share received data amongst themselves to infer sensitive information, such as observation data or cluster labels of the honest nodes.

\subsection{Main Requirements}\label{problemSetup}

A trustworthy, distributed \(k\)-means clustering approach for ECS must meet the following two core requirements:

1) Clustering correctness: The clustering result produced by the trustworthy \(k\)-means algorithm should be identical to that of a traditional \(k\)-means algorithm without trust concerns.

2) Individual data protection: The sensitive data of each honest node—specifically, the observation data and final cluster label—must remain protected throughout the algorithm's execution under a passive adversary model.

\section{Building Block}\label{prelimilary}

In this paper, we opt for the additive secret sharing approach \cite{gupta2017privacy} due to its simplicity. Below, we introduce the additive secret sharing algorithm, which will later serve as a foundational component of our proposed method.
 
The purpose of the adopted algorithm is to securely compute the global average of each node’s input, denoted as \( a_i \), across the network while keeping individual inputs confidential. This algorithm ensures both accurate averaging and complete trustiness, provided that each honest node has at least one honest neighboring node.

The algorithm comprises three primary steps. First, each node obfuscates its input \( a_i \) by generating an additive secret share \( s_i^i \), following the principles of additive secret sharing \cite{Cramer2015}. This step is critical for preserving privacy, as the obfuscation ensures that the distribution of \( s_i^i \) is statistically independent of the original input, providing perfect trustiness.

In the second step, since the obfuscated shares \( s_i^i \) no longer reveal sensitive information, they can be safely used in any distributed averaging protocol to compute the global average, denoted as \( \bar{s} \), across all nodes \( i \in V \). 

Finally, the true average of the original inputs is reconstructed from the obfuscated average via:
\[
s_{\text{a}} = \frac{1}{n} \left( \bar{s} \times n \mod p \right),
\]
where \( p \) is a sufficiently large prime that satisfies \( p > \sum_{i \in V} a_i \). Algorithm \ref{proAdditive} outlines the steps in detail, where \( p \) is chosen to exceed \( \sum_{i \in V} a_i \), and \( F \) represents the set of integers modulo \( p \).

\begin{algorithm}
\caption{Trustworthy Distributed Average Consensus using Additive Secret Sharing}
\label{proAdditive}
\textbf{Step 1: Additive Randomization}

\begin{algorithmic}[1]
\STATE Initialize each node $i \in V$ with input $a_i$.
\WHILE{Node $i$ has active neighbors $k \in d_i$}
    \STATE Generate a random value $r_i^k$ uniformly from $F$.
    \STATE Send $r_i^k$ to neighboring node $k$.
\ENDWHILE

\FOR{each neighboring node $k \in d_i$}
    \STATE Compute obfuscated share $s_i^k = (r_i^k - r_k^i) \mod p$.
\ENDFOR

\STATE Compute private share for node $i$:
\[
    s_i^i = \left( a_i - \sum_{k \in d_i} s_i^k \right) \mod p.
\]

\textbf{Step 2: Distributed Averaging}
\WHILE{Distributed averaging protocol is running}
    \STATE Use $s_i^i$ as input to compute the global obfuscated average $\bar{s}$ via protocols such as \cite{johansson2008faster, boyd2006randomized, dimakis2010gossip}.
\ENDWHILE

\textbf{Step 3: Average Consensus Computation}
\STATE Each node computes the final average as:
\[
    s_{\text{a}} = \frac{1}{n} \left( \bar{s} \times n \mod p \right).
\]
\end{algorithmic}
\end{algorithm}

% \begin{algorithm}
% %\vskip -1pt
% \caption{Trustworthy Distributed Average Consensus using Additive Secret Sharing}

% \textbf{Step 1: Additive Randomization}

% \begin{algorithmic}[1] \label{proAdditive}
% \STATE Each node \( i \in V \) holds an input \( a_i \) and generates a set of random values \( \{ r_i^k \mid k \in d_i \} \) uniformly from \( F \). Each \( r_i^k \) is sent to the corresponding neighboring node \( k \).
% \STATE Node \( i \) computes a set of obfuscated shares \( \{ s_i^k = (r_i^k - r_k^i) \mod p \mid k \in d_i \} \).
% \STATE Node \( i \) then constructs its private share \( s_i^i \) based on its input \( a_i \) and the obfuscated shares from its neighbors:
% \[
% s_i^i = \left( a_i - \sum_{k \in d_i} s_i^k \right) \mod p.
% \]
% \textbf{Step 2: Distributed Averaging}
% \STATE Each node \( i \in V \) uses its private share \( s_i^i \) as input to a distributed averaging protocol \cite{johansson2008faster, boyd2006randomized, dimakis2010gossip} to compute the global obfuscated average \( \bar{s} \).

% \textbf{Step 3: Average Consensus Computation}
% \STATE Each node computes the final average as
% \[
% s_{\text{a}} = \frac{1}{n} \left( \bar{s} \times n \mod p \right)
% \]
% .
% \end{algorithmic}
% \vskip -2pt
% \end{algorithm}

\section{Proposed Approach}\label{proposed}
The proposed approach includes two parts: cluster assignment and privacy-preserving cluster center updating. Firstly, each node $i$ computes its cluster label $l_{i}(t)$ based on \eqref{clusterLabel}, this local computation will not violate the privacy as it does not require any information sharing. Secondly, a privacy-preserving cluster center updating algorithm is proposed. When updating the cluster centers in ECS, the information exchange required in the distributed data aggregation protocols  will inevitably reveal the private data held by each node. Therefore, we apply the abovementioned privacy-preserving distributed averaging algorithm to protect the private data held by each node.  Note that even we aim to protect the final cluster label $l_i^{f}$  from revealing, the intermediate cluster label $l_i(t)$  should also be protected as secret since it is highly correlated with the final cluster label especially during the last few iterations. As a consequence, the goal is to protect both the  observation data and intermediate cluster label in cluster center updating step. To do so, each node first constructs the extended  observation data and index vector:
\begin{align}\label{extendVec}
\textbf y_{i}^{l_{i}}(t)=
  \begin{bmatrix}
    0,\dots, x_{i},\dots, 0
  \end{bmatrix}, \nonumber\\
  \textbf e_{i}^{l_{i}}(t)=
  \begin{bmatrix}
    0,\dots,  1 ,\dots,0
  \end{bmatrix}, 
\end{align}
where $\textbf y_{i}^{l_{i}}(t)\in \mathbb{R}^{d\times k} $ has the  observation data $ x_{i}$ in the $l_i(t)$-th column and zeros in elsewhere. Similarly,  the extended index vector $\textbf e_{i}^{l_{i}}(t)\in \mathbb{R}^{k}$ has its $l_i(t)$-th entry as one and zeros elsewhere.  Note that inserting zeros in $\textbf y_{i}^{l_{i}}(t)$ and $\textbf e_{i}^{l_{i}}(t)$ is to keep the sum of  observation data and number of nodes in each cluster unchanged.

After that, apply Algorithm \ref{proAdditive} to securely compute the averages on the constructed $\textbf y_{i}^{l_{i}}(t)$ and $\textbf e_{i}^{l_{i}}(t)$ over the network, respectively. The computed averages would be normalized sum and node number in each cluster:  
\begin{align}\label{ave:YN}
    \bar{\textbf Y}(t)=\frac{1}{n}\sum_{i \in V} \textbf y_{i}^{l_{i}}(t),\nonumber\\
    \bar{\textbf N}(t)=\frac{1}{n}\sum_{i \in V} \textbf e_{i}^{l_i}(t). 
\end{align}
As a consequence, the cluster centers can be updated as:         
\begin{align}\label{aveResult}
\textbf c^{p}_{j}(t+1)=\frac{\bar{\textbf Y}_{j}(t)}{\bar{\textbf N}_{j}(t)}, j \in K.
\end{align}

Details of the proposed approach are summarized in Algorithm \ref{proAlgorithm}.
\begin{algorithm}
\vskip -2pt
\caption{Proposed approach}
\begin{algorithmic}[1] \label{proAlgorithm}
%\STATE Predefine the cluster number as $K$ and broadcast over the whole network.
\STATE Randomly initialize the cluster centers as $ c_{j}(0), j \in K$.
\STATE For iteration $t=0,1, 2, 3, ..., T$
\\
%\vspace{1.2mm}
\textbf{Cluster assignment:~~~~}
\STATE Each node $i$ computes its cluster label $l_i(t)$ based on \eqref{clusterLabel}.
\\
%\vspace{1.2mm}
\textbf{Privacy-preserving cluster center updating:}
\STATE Each node $i$ constructs its extended private $\textbf y_{i}^{l_{i}}(t)$ and extended index vector $\textbf e_{i}^{l_i} $ based on \eqref{extendVec}.
\STATE Each node $i$ set $\textbf y_{i}^{l_{i}}(t)$ and $\textbf e_{i}^{l_i}(t)$ as the inputs (i.e., $a_{i}$) in Algorithm \ref{proAdditive} to compute the average result  $\bar{\textbf Y}(t)$ and $\bar{\textbf N}(t)$ over the network, respectively.
\STATE Each node update the cluster centers based on \eqref{aveResult}.
\STATE Repeat until Convergence (Clustering Center does not change anymore).
\STATE End

\end{algorithmic}
\vskip -2pt
\end{algorithm}

\section{Analysis}
In this section, we will analyze the proposed approach in terms of clustering accuracy and privacy.

\subsection{Clustering accuracy analysis}
By inspecting \eqref{ave:YN} and \eqref{aveResult}, we can see that the cluster centers computed by the proposed approach are identical to the original one shown in \eqref{clusterLabel}:
 \begin{align*}
 c^{p}_{j}(t+1)
 =\frac{\bar{\textbf Y}_{j}(t)}{\bar{\textbf N}_{j}(t)}=\frac{1}{N_{j}} \sum_{ x_{i} \in {\cal C}_{j(t)} }  x_{i}= c_{j}(t+1), j \in K.
\end{align*}
Hence, the proposed approach will output identical cluster centers to the conventional algorithms thereby having no tradeoff in clustering performance and privacy-preservation. 

\subsection{Privacy Analysis}\label{sectionAnalysis}
As explained before, the cluster assignment step does not violate the privacy concern. Therefore, we will focus on the privacy analysis of the cluster center updating step.  
%Throughout the whole algorithm execution,  the public known information includes the cluster centers $\textbf c_{j}(t), j \in K$, the average of extended  observation data $\bar{\textbf Y}(t)$ and index vector $\bar{\textbf N}(t)$. The cluster centers can be computed from $\bar{\textbf Y}(t)$ and $\bar{\textbf N}(t)$, and we, therefore, only need to consider the information leakage cause by $\bar{\textbf Y}(t)$ and $\bar{\textbf N}(t)$ under passive attack. 

Let \( V_{c} \) and \( V_{h} \) represent the sets of passively corrupted nodes and honest nodes, respectively. To simplify the analysis, we assume initially that the set of honest nodes \( V_{h} \) forms a connected subgraph. As shown in Algorithm \ref{proAdditive}, the input of any honest node \( i \in V_{h} \) remains secure as long as \( |N_{i} \cap V_{h}| \neq \emptyset \), recall \( N_{i} \) denotes the neighbors of node \( i \). In this case, the only information disclosed is the aggregated sum of the inputs from all honest nodes. This ensures that no individual input from any honest node is directly exposed, thereby preserving privacy. 
%This implies the following: for each isolated iteration, each honest node's  observation data and intermediate cluster label is perfectly protected as long as it has one honest neighbour.  

Since k-means clustering is an iterative process where the successive iterations are highly correlated. Therefore, we will first analyze the privacy in each iteration and then combine the information in all iterations. 
%Without loss of generality, assume the corrupted nodes attempt to infer the private data $x_i$.  ${I}(t)$ denotes the information obtained by the corrupted nodes is To quantitatively measure the individual privacy, we 
%and there are some additional information for example the sum of honest nodes' private data is known by the corrupted nodes. We will therefore first analyze the privacy in each isolated iteration and then combine information in all the iterations.  

\subsubsection{Privacy analysis in each iteration}
At each iteration $t$, denote  ${I}(t)$ as the information set obtained by the corrupted node set $V_{c}$ at current iteration. As the sum of honest nodes' inputs as it can be deduced from the mean of each cluster and the inputs of the corrupted nodes, we have 
\begin{align}
\mathcal {I}(t)= \{\textbf s_{y}^{H}(t)=\sum_{i\in\mathcal {H}} \textbf y_{i}^{l_{i}}(t),
\textbf s_{e}^{H}(t)=\sum_{i\in\mathcal {H}} \textbf e_{i}^{l_i}(t)\}.
\end{align}
Let $k$ subsets $\bigcup_{j} \Lambda_{j}(t)=V_{h}$ denotes the set of nodes in each cluster and there are $n_{j}(t)$ nodes in $\Lambda_{j}(t)$, the total number of honest nodes is $n_{h}=\sum_{j \in K } n_{j}(t)$. The information set is then $\mathcal {I}(t)= \{\textbf y_{j}(t)=\sum_{i \in \Lambda_{j}(t) } \textbf x_{i}| j \in K \}$.
Define all the  observation data $ x_{i}$ held by honest nodes $i$ as random variable $X_{i},i \in V_{h}$, and $\sum_{i \in \Lambda_{j}(t) }  x_{i}$ in each cluster as random variable $Y_{j}(t)$. Assume the  observation data $ x_{i}$
held by each node is independent with each other, therefore the random variable $Y_{j}(t)$ can be seen as the sum of a number of independent random variables as:
$Y_{j}(t)=\sum_{i \in \Lambda_{j}(t) } X_{i}, j \in K$.
Thus the information set $\mathcal {I}(t)$ contains the joint random variables $\{Y_{1}(t),Y_{2}(t),...,Y_{k}(t)\}$, and they are also independent with each other.
Without loss of generality, for any individual honest node $i\in V_{h}$ with differential entropy $h(X_{i})$, our goal is to compute the information leakage regarding to the individual node $i$ based on the information $\mathcal {I}(t)$ obtained by the corrupted nodes, which is the mutual information $I(X_{i};Y_{1}(t),...,Y_{k}(t))$. By chain rule
\begin{align}\label{mutualAll}
I(X_{i};Y_{1}(t),...,Y_{k}(t))=&\sum_{j\in K} (I(X_{i};Y_{j}(t)|Y_{j-1}(t),...,Y_{1}(t)))\nonumber\\
=&\sum_{j\in K} (I(X_{i};Y_{j}(t))), 
%\\
%=&\sum_{j\in K} \frac{n_{j}(t)}{n_{h}} (h(Y_{j}(t))-h(Y_{j}(t)-X_{i}))
\end{align}
 note that there are at least $k-1$ zeros in $I(X_{i};Y_{j}(t)),j \in K$ since $X_{i}$ can only locate in one cluster. For example, if $X_{i}$ is located in cluster 1, then all other mutual information $I(X_{i};Y_{j}(t)),j \neq 1 $ are zeros since $X_{i}$ and $Y_{j}(t)$ are independent with each other. We then have
\begin{align} \label{infoLeak}
&I(X_{i};Y_{j}(t)) \nonumber\\
=&\int_{x_{i}}\int_{y_{j}(t)}p(x_{i},y_{j}(t))log \frac{p(x_{i},y_{j}(t))}{p(x_{i})p(y_{j}(t))}\nonumber\\
=&\sum_{u=1}^{k}Pr\{x_i \in \Lambda_{u}(t)\}\int_{x_{i}}\int_{y_{j}(t)} p(x_{i},y_{j}(t)|x_i \in \Lambda_{u}(t))\nonumber\\
& log \frac{p(x_{i},y_{j}(t))}{p(x_{i})p(y_{j}(t))}\nonumber\\
%=&\sum_{u=1}^{k}Pr\{x_i \in \Lambda_{u}(t)\} I(X_{i};Y_{u}(t)) \nonumber\\
=&Pr\{x_i \in \Lambda_{j}(t)\} I(X_{i};Y_{j}(t)|x_i \in \Lambda_{j}(t))) \nonumber\\
= &  \frac{n_{j}(t)}{n_{h}}I(X_{i};Y_{j}(t)|x_i \in \Lambda_{j}(t)) \nonumber\\
=&
 \begin{cases}
0, n_{j}(t) > 1\\
 \frac{1}{n_{h}}h(X_{i}), n_{j}(t) = 1,
\end{cases}
\end{align}

where $p(x_{i},y_{j}(t)|x_i \in \Lambda_{u}(t)) =p(x_{i},y_{u}(t)) \sigma_{ju}$ and  $ \sigma_{ju}=  
 \begin{cases}
0, j \neq u\\
 1, j=u,
\end{cases} $ is the indicator function denoting if node $i$ belongs to cluster $j$. 
 $\frac{n_{j}(t)}{n_{h}}=Pr\{x_i \in \Lambda_{j}(t)\}$ is the probability when node $i$ belongs to cluster $j$.
We can see that mutual information is zero if there are more than one node in certain cluster as the perfect security is guaranteed. If there is only one honest node within certain cluster, the mutual information is $\frac{1 }{n_{h}} h(X_{i})$ since  $I(X_{i};Y_{j}(t)=X_{i}|x_i \in \Lambda_{j}(t))=h(X_{i})$. 
\\
\subsubsection{Privacy analysis across all the iterations}
Here, we will now analyze whether a clever combination of information across different iterations will increase the information leakage. The accumulated information $\mathcal {I}$ obtained by the corrupted nodes across all the iterations is given by $\mathcal {I}=\{\mathcal {I}(t)|t=1,2,...,T\}.$
We can derive some extra information by comparing the information obtained in any two iterations $m,k \in {1,2,...,T}$:
%\begin{align}
$\mathcal {I}_{mk}=\{\textbf s_{y}^{H}(m)-\textbf s_{y}^{H}(k),\textbf s_{e}^{H}(m)-\textbf s_{e}^{H}(k)\}$.
%\end{align}
Assuming the worst case, namely that only one node moves from one cluster to another within two different iterations, then the private value of this node can be computed. Repeat this process, we can see that the accumulated information in the most malicious case would contain a collection of all  observation data held by each honest node, i.e., $\mathcal {I}=\{X_{i}| i \in V_{h}\}$. The probability of inferring any specified $X_{i}$ held by node $i$ based on $\mathcal {I}$ would be $\frac{1}{n_{h}}$. Similar as (\ref{mutualAll}) and (\ref{infoLeak}) , the maximum mutual information can thus be computed as   

\begin{align}
I(X_{i};X_{1},...,X_{i},...)= \frac{1}{n_{h}}h(X_{i}),
\end{align}
%\begin{align*}
%I(X_{i};\mathcal {I})&=
%I(X_{i};X_{1},...,X_{i},...)
%&=\frac{1}{n_{h}}I(X_{i};X_{i})+\sum_{m \in V_{h}, m \neq i } I(X_{i};X_{m})\\
%&= \frac{1}{n_{h}}h(X_{i}).
%\end{align*}
\\
Combined with the analysis within each iteration, we can conclude that the maximum information leakage about the  observation data $ x_{i}$ throughout the whole algorithm execution is $\frac{1}{n_{h}} h(X_{i})$. Note that if the honest node set $V_{h}$ is not connected, then it can be divided into several connected subsets $\bigcup_{m} V_{h_m}=V_{h}$, the privacy analysis is exactly the same as above by considering each connected subset separately. 

%As for the information leakage of final cluster label $l_{i}^{f}$, this final cluster label is determined if and only if there is only one non-zero column in both $ s_{y}^{H}(T)$ and $\textbf s_{e}^{H}(T)$ after convergence. 

Privacy preservation in this framework is ensured under the condition of an honest majority, as is similarly assumed in \cite{lin2005privacy}. This assumption means that more than half of the nodes in the network remain honest and do not collude to compromise privacy. Consider an honest node \(i\) within a connected subset \(V_{h_m}\) consisting of \(M\) nodes. Based on this setting, the following observations can be made regarding privacy:

\begin{enumerate}
    \item Final cluster label \(l_i^f\): The privacy of the final cluster label is effectively preserved under the honest majority assumption. Specifically, \(l_i^f\) can only be revealed if all \(M\) honest nodes within the subset \(V_{h_m}\) end up in the same cluster. However, this scenario is highly improbable under the honest majority condition. For instance, assuming each node is equally likely to be assigned to any of the \(k\) clusters, the probability of all \(M\) nodes being in the same cluster is \((\frac{1}{k})^M\). As \(M\) increases, this probability diminishes exponentially, making it exceedingly unlikely in practical settings.
    \item Private data \(x_i\): The privacy of the observation data \(x_i\) is similarly safeguarded. The information leakage associated with \(x_i\) is upper-bounded by \(\frac{1}{M} h(X_i)\). This upper bound indicates that as the number of honest nodes \(M\) in the connected subset increases, the leakage of information about \(x_i\) decreases. Under the honest majority condition, \(M\) is typically large, ensuring that the actual information leakage is minimal. This relationship highlights the importance of maintaining a robust honest majority, as it significantly mitigates privacy risks.
\end{enumerate}

%\section{Experimental validations}
\section{Comparisons}\label{sectionCompare}
In this section, we compare the proposed approach with the existing user-centric approach in terms of several parameters shown in Table \ref{table1}. 

We can see that both approaches consider the same distributed setting adversary model. The user-centric approach makes use of encryption techniques like Paillier cryptosystems and secure function evaluation adopted in \cite{sakuma2010large}, which usually require expensive exponential functions. As a consequence, the required computational complexity and communication bandwidth are much higher than for the proposed approach. For example, if a signal sample usually take 8-16 bits but its encrypted counterparts are usually 1024 or more bits long \cite{lagendijk2013encrypted}. In addition, an extra trusted third party is required in the user-centric approach, something that is usually not practical in applications like ECS. Moreover, the achieved perfect  (i.e., information-theoretic)  security model in the proposed approach is stronger than the computational security model, where the former can guarantee the security against the adversary equipped with unlimited computational power while the later one can not (see \cite[Section 1.3.1]{Cramer2015} for further details). 
%the computational security, which is based on the assumption that the adversary is computationally limited, perfect (i.e., information-theoretic) security model is considered in the proposed approach, even with an adversary with unlimited computational resources (see \cite[Section 1.3.1]{Cramer2015} for further details). 
As shown in the distributed averaging step of Algorithm \ref{proAdditive}, the proposed approach is quite general and can adopt all distributed averaging algorithms for the cluster center updating step while only random gossip algorithm is used in \cite{sakuma2010large}.
Concerning privacy, the cluster centers are not considered private information in the proposed approach as they only reflect group property owned by all the individuals within each cluster. The proposed approach has an assumption of honest majority, while this assumption is not required in \cite{sakuma2010large}.

\begin{table}[t]
\begin{center}
\centering \caption{Comparison of the Proposed and User-Centric Approaches} \label{table1}
\scriptsize
\begin{tabular}{lll}
\toprule
%\multicolumn{1}{c|}{ } & \multicolumn{2}{c}{Approaches}\\

\textbf {} & \textbf {Proposed} & \textbf {User-centric approach} 

\cite{sakuma2010large}\\

\midrule
Adversary Model&Passive&Passive\\
\hline
Graph Topology&Fully distributed&Fully distributed\\
\hline
Involved Function&Linear &Exponential\\
\hline
Signal Bit Length&$8-16$ &$512/1024$ \\
\hline
Trusted Third Party&No&Yes\\
\hline
Security Model& Perfect &Computational\\
\hline
Distributed Algorithm&Arbitrary \cite{johansson2008faster,boyd2006randomized,
dimakis2010gossip,boyd2011distributed,zhang2018distributed} &Gossip \cite{boyd2006randomized}\\
\hline

Privacy Concerns& $ x_{i},l_{i}(t)$ &$ x_{i},l_{i}(t), c_{j}(t)$\\

\bottomrule
\end{tabular}
\end{center}
\vskip -16pt
\end{table}

\vspace{-5mm}

\section{Conclusions and Future Work}
\label{Sect-Conclu}

In this paper, we presented a lightweight, trustworthy, and fully distributed k-means clustering algorithm tailored for ECS, such as sensor networks. By incorporating the additive secret sharing scheme during the cluster center update process, the proposed algorithm not only achieves identical clustering results to the original k-means but also significantly reduces computational complexity and communication bandwidth. These characteristics make it ideal for the resource-constrained and latency-sensitive environments of ECS.

The method’s accuracy and Trustworthiness remain valid as long as the honest majority assumption holds, which is generally reasonable in the context of ECS. However, to enhance the algorithm’s robustness in less cooperative or more adversarial scenarios, future work will explore strategies to relax or eliminate the reliance on the honest majority assumption.

In summary, the proposed lightweight approach effectively addresses the performance and trustworthiness requirements of edge computing systems. We also plan to validate the proposed approach on real-world edge computing datasets and platforms to further demonstrate its practical applicability and scalability.

%\pagebreak

%\bibliographystyle{IEEEtran}
%%\bibliography{shamir}
%\bibliographystyle{package_style/IEEEbib}
%\bibliography{refs/IEEEabrv,refs/myabrv,refs/shamir}
%\bibliographystyle{package_style/IEEEbib}
%\bibliography{refs/IEEEabrv,refs/myabrv,refs/dualpath}

\bibliographystyle{IEEEtran}

\bibliography{dualpath}

\end{document}